\documentclass[12pt,a4paper]{article}
\usepackage{latexsym,amsfonts,graphicx}

\newtheorem{thm}{Theorem}[section]
\newtheorem{lem}{Lemma}[section]
\newtheorem{cor}{Corollary}[section]
\newtheorem{prop}{Proposition}[section]

\newcommand{\1}{\mathbb{I}}
\newcommand{\C}{\mathbb{C}}
\newcommand{\T}{\mathbb{T}}
\newcommand{\R}{\mathbb{R}}

\newcommand{\Dom}{\mbox{Dom}}
\newcommand{\cL}{{\cal L}}

\newcommand{\Th}{\Theta}
\newcommand{\Ph}{\Phi}

\newcommand{\Js}[1]{F_{+{#1}}}
\newcommand{\Jsb}[1]{F_{-{#1}}}
\newcommand{\Jsd}[1]{F_{\pm{#1}}}
\newcommand{\Jsa}[1]{F^{\dagger}_{-{#1}}}

\newcommand{\Jsad}[1]{F^{\dagger}_{\pm{#1}}}

\newcommand{\Jf}[1]{{\sf F}_{+{#1}}}
\newcommand{\Jfb}[1]{{\sf F}_{-{#1}}}
\newcommand{\Jfd}[1]{{\sf F}_{\pm{#1}}}

\newcommand{\Jfa}[1]{{\sf F}^{\dagger}_{-{#1}}}
\newcommand{\Jfab}[1]{{\sf F}^{\dagger}_{+{#1}}}
\newcommand{\Jfad}[1]{{\sf F}^{\dagger}_{\pm{#1}}}

\newcommand{\M}[2]{M^{#1}_{+{#2}}}
\newcommand{\Mb}[2]{M^{#1}_{-{#2}}}
\newcommand{\Md}[2]{M^{#1}_{\pm{#2}}}

\newcommand{\Ma}[2]{M^{\dagger\,{#1}}_{-{#2}}}
\newcommand{\Mab}[2]{{M}^{\dagger\,{#1}}_{+{#2}}}

\newcommand{\Sw}{\Psi}
\newcommand{\Swa}{\Psi^{\dagger}}
\newcommand{\eSw}{\Xi \rule{0mm}{4.5mm}}
\newcommand{\eSwa}{\Xi^{\dagger} \rule{0mm}{4.5mm}}

\newcommand{\Nb}[1]{N_{-{#1}}}

\newcommand{\Nab}[1]{N^{\dagger}_{+{#1}}}
\newcommand{\Oab}[1]{O^{\dagger}_{+{#1}}}

\title{Inverse Scattering on Matrices with Boundary Conditions}
\author{Mark Harmer}

\begin{document}
\maketitle 
\begin{abstract}
We describe inverse scattering for the matrix Schr\"{o}dinger operator with general selfadjoint boundary conditions at the origin using the Marchenko equation. Our approach allows the recovery of the potential as well as the boundary conditions. It is easily specialised to inverse scattering on star-shaped graphs with boundary conditions at the node. 
\end{abstract}

\section{Introduction}
The monograph by Agranovich and Marchenko \cite{Agr:Mar} provides {\em the} description of the inverse scattering problem for the matrix Schr\"{o}dinger operator on the semi-axis with Dirichlet boundary conditions at the origin. This followed pioneering work by Gelfand, Jost, Krein, Levitan, Marchenko and other authors on the inverse problem. For a more complete description of the history of this problem see the series of papers by Faddeev \cite{Fad1,Fad2}. \\
In this paper we generalise the work of Agranovich and Marchenko. We consider the matrix Schr\"{o}dinger operator with general selfadjoint boundary conditions at the origin. As a consequence the boundary conditions, in the form of a unitary matrix $U$, appear as a crucial element in our discussion of the direct and inverse problems. In particular we see that the scattering data, due to the continuous part of the spectrum, is of the form $S(k) - \hat{U}$ where $S(k)$ is the scattering matrix and $\hat{U}$ is a matrix directly related to the boundary conditions (in fact it is the asymptotic of the scattering matrix). From our discussion of the inverse problem it is easy to see that the boundary conditions $U$ along with the potential matrix are recovered in the course of the solution. \\
The consideration of general boundary conditions, instead of Dirichlet boundary conditions, increases the complexity of the problem from an analytic and algebraic perspective. From an analytic point of view the characterisation of the set of scattering data becomes far more difficult. We are far from finding an equivalence between the set of permissible matrix Schr\"{o}dinger operators and a set of scattering data (on the other hand a complete characterisation is given in \cite{Agr:Mar} for Dirichlet boundary conditions). In this paper we have concentrated mainly on the algebraic complications that arise in the consideration of general boundary conditions. \\
There has been a lot of work on systems with singular or finite rank perturbations \cite{AGHH,Alb:Kur,Pav} and in particular they play an important role in the study of operators on graphs \cite{Bul:Tre,Car1,Car2,Car3,Exn,Exn:Seb,Gera:Pav,Har4}. Such systems may be useful in modeling the behaviour of real physical problems, in particular nanoelectronic devices \cite{Exn:Nem,ESS,Kost:Sch,Mel:Pav,MPPRY,Har6}. In these studies the choice of finite rank perturbation, or boundary condition, is an important part of the modeling of the dynamical system. On the other hand, as far as this author is aware, no work has been done on the inverse problem where the selfadjoint boundary conditions play a central role and are recovered as part of the solution. It is for this reason we believe that this paper will be of interest. \\
As mentioned above many of the papers using finite rank perturbations in applications consider operators on graphs, rather than, as in our case, matrix systems. However, it is easy to see that the matrix system is a generalisation of the star-shaped graph which has an important, even central, role in applications (see \cite{BMPY} where there is a discussion of modeling general quantum networks using such a system). \\
There are many other approaches to the inverse problem aside from the approach we have adopted here. There is an abundance of  inverse spectral methods \cite{Aro:Dym,Rem,Sim} and the boundary control method \cite{ALP,Avd:Iva,Bel1,Bel2} is another very useful approach. The boundary control method is particularly powerful when applied to inverse problems on finite graphs as it allows the recovery of the operator on the graph as well as the geometry of the graph \cite{Bel2}! The matrix Schr\"{o}dinger operator is easily seen to generalise the Schr\"{o}dinger operator on a graph with $n$ rays and in \cite{Gera,Kur:Ste} the authors consider the inverse problem on graphs using similar methods to the method used here (we note however that there are some errors in \cite{Gera}, specifically the inverse problem is not correctly described). \\
The contents of this paper are based on the results of the PhD thesis \cite{Har}.
\section{The Matrix Schr\"{o}dinger Operator with Selfadjoint Boundary Conditions}
Let us consider the Hilbert space $L^2 (\R_+ ; \C^n)$ consisting of the set of functions from $\R_+ \equiv
[0,\infty )$ to $\C^n$ satisfying
$$
\left\| f\right\|^{2} \equiv \int^{\infty}_{0} | f(x) |^{2}\, dx < \infty 
$$
with associated inner product
$$
\langle f,g\rangle = \int^{\infty}_0 f(x)^* \cdot g(x)\, dx \, .
$$
Here $|\cdot |$ is the usual norm in $\C^n$ and $\cdot^*$ denotes the complex conjugate transpose for vectors and matrices. \\
We define the matrix Schr\"{o}dinger operator
$$
\cL_0 \equiv -\frac{d^2}{dx^2} + Q(x) 
$$
where $Q=Q^*$ is a hermitian matrix which satisfies
\begin{equation}\label{bcndxx}
\int^{\infty}_0 (1 + t) \vert Q(t) \vert\, dt < \infty\, .
\end{equation}
The norm, $\vert A\vert$, of a matrix acting on $\C^n$ is defined here as the maximal eigenvalue of the matrix. The domain of $\cL_0$ is chosen to be the set of smooth functions with the (closed) support a compact subset of $(0,\infty )$
$$
\Dom (\cL_{0})= C^{\infty}_{0}(\R_+ ;\C^n )\, .
$$
In this case $\cL_0$ has deficiency indices $(n,n)$---we are in the limit point case---and the selfadjoint extensions of $\cL_0$ can be parameterised by an $n\times n$ unitary matrix $U$ using Neumann extension theory \cite{Akh:Glz}. Here we take a slightly different but equivalent approach \cite{Har4} whereby the selfadjoint extensions are described by the boundary conditions at the origin
\begin{equation}\label{basicbc}
\left. \frac{i}{2}(U^{*}-\1 )\cdot f \right|_0 +
\left. \frac{1}{2}(U^{*} +\1 )\cdot f_x \right|_0 = 0
\end{equation}
for some unitary $U$ which is fixed in the remainder of the discussion. The latin subscript denotes differentiation. \\
In order to simplify the presentation we assume that there are no virtual levels---ie. zero is not an eigenvalue---of $\cL$.
\section{Properties of the Solutions}
Here we consider solutions $Y(x,k)$ of the eigenvalue equation 
\begin{equation}\label{be}
\cL\, Y = k^2\, Y
\end{equation} 
for the most part ignoring boundary conditions (\ref{be}) and square integrability.
It is convenient to define the involution
$$
Y^{\dagger} (x,k)\equiv Y^{*}(x,\bar{k})  
$$
as it is then clear that the Wronskian
$$
W\{\Phi^{\dagger} , \Psi \} = \Phi^{\dagger} \Psi_x - \Phi^{\dagger}_x \Psi \, ,
$$
of two solutions of (\ref{be}) at the same value of $k$ are constant (note the Wronskian as a function of $\Phi$ and $\Psi$ is a matrix valued hermitian symplectic form \cite{Har4}. We also note that if we swop the order $W\{\Phi , \Psi^{\dagger} \}$ is {\em not} constant). \\
We define the standard solutions:
\begin{thm}\label{thss}
There exist solutions, $\Th$ and $\Ph$, of the matrix equation
(\ref{be}) subject to the condition (\ref{bcndxx}) which
satisfy the following boundary conditions at the origin
\begin{eqnarray*}
& \lim_{x\rightarrow 0} \Ph (x,k) = 0\, , \hspace{5mm} 
 \lim_{x\rightarrow 0} \Ph_x (x,k) = \1 & \\ \vspace{5mm}
& \lim_{x\rightarrow 0} \Th (x,k) = \1\, , \hspace{5mm} 
 \lim_{x\rightarrow 0} \Th_x (x,k) = 0 & 
\end{eqnarray*}
and are {\em entire} functions in the variable $k$.
\end{thm}
The proof of this theorem follows a standard argument (see theorem 1.2.1
of Agranovich and Marchenko \cite{Agr:Mar} for $\Ph$, the discussion for $\Th$ is similar \cite{Har}). We define the Jost solution using the asymptotic boundary condition
$$ 
\lim_{x\rightarrow\infty} F (x,k) =e^{ ikx}\1 \, .
$$
\begin{thm}\label{thJsi}
The Jost solution $F (x,k)$ and its derivative $F_x (x,k)$ are analytic
in the upper half-plane of the variable $k$ and continuous there, up to and 
including the real axis.
\end{thm}
This too is a standard result (it is a slight extension of theorem 1.3.1 in \cite{Agr:Mar}, see \cite{Har} for details). As a notational convenience we define
$$
\Jsd{}(x,k) \equiv F (x,\pm k) \, .
$$
\begin{thm}\label{thJs}
The Jost\index{Js} solution $\Js{}(x,k)$ can be written in terms of the
transformation operator
\begin{equation}\label{beJT}
\Js{} (x,k) = e^{ikx}\1 + \int^{\infty}_{x} K(x,t) e^{ikt} dt
\end{equation}
with kernel $K(x,t)$ which is bounded, absolutely integrable with respect to its second argument
and the derivative $K_x (x,t)$ is also absolutely integrable with respect to
its second argument. The kernel and the potential matrix are related by
\begin{equation}\label{beIP}
\mbox{} -2 \frac{d K(x,x)}{dx} = Q(x) \, .
\end{equation}
\end{thm}
Again the proof may be found in \cite{Agr:Mar}, see mainly theorem 1.3.1. \\
Finally we would like to define the scattered waves. We first define the
function $\eSw (x,k)$ as the solution of (\ref{be}) with boundary conditions
\begin{equation}\label{eSwo}
\left. \eSw \right|_0 = \frac{1}{2}( U + \1 ) \equiv A \, ,\hspace{5mm} 
\left. \eSw_x \right|_0 = \frac{i}{2}( U -\1 ) \equiv B\, . 
\end{equation}
Here $U$ is the fixed unitary matrix which defines the selfadjoint boundary conditions; $A$ and $B$ are defined above as a convenient shorthand. It is easy to see that $A^* B = B^* A$ from which we immediately get that $\eSw (x,k)$ satisfies the selfadjoint boundary conditions (\ref{basicbc}). Clearly we can write $\eSw$ in terms of the standard solutions
$$
\eSw (x,k) = \Th (x,k) A + \Ph (x,k) B\, , 
$$
from which we see that $\eSw$ is entire in $k$, or in terms of the Jost solutions
$$
\eSw (x,k) = \Jsb{}(x,k)\Mb{}{}(k) + \Js{}(x,k)\M{}{}(k) \, ,
$$
where $\Md{}{}$ are some functions of the spectral parameter. Taking the Wronskian
\begin{eqnarray*}
W\{ \Jsad{}, \eSw \} & = & \left.\left[ \Jsad{} \eSw_x - \Jsad{,x} \eSw \right]\right|_0 
= \Jfad{} B - \Jfad{,x} A \\
& = & \pm 2ik \Md{}{}
\end{eqnarray*}
we get an expression for $\Md{}{}$ in terms of the Jost solutions. Here we use the constancy of the Wronskian and the two expressions for $\eSw$ in turn. The functions
$$
\Jfd{} (k) = \Jsd{} (0,k)  \, ,\hspace{5mm} \Jfd{,x} (k) = \Jsd{,x} (0,k)
$$
are known as the Jost functions. This gives us 
\begin{equation}\label{sdmi}
\Md{}{} = \pm\frac{1}{2ik}\left[ \Jfad{} B - \Jfad{,x} A \right] \, .
\end{equation}
Choosing appropriate $U$ we can make $\eSw$ equal the standard solutions. This allows us to write the standard solutions in terms of the Jost solutions. Evaluating at the origin we get the following identities for the Jost functions
\begin{eqnarray}
\Jfb{} \Jfa{,x} - \Jf{} \Jfab{,x} & = & 2ik\1 \label{PsWi} \\ 
\Jf{,x} \Jfab{} - \Jfb{,x} \Jfa{} & = & 2ik\1 \label{PsWii} \\
\Jfb{,x} \Jfa{,x} - \Jf{,x} \Jfab{,x} & = & 0 \label{PsWiii} \\ 
\Jfb{} \Jfa{} - \Jf{} \Jfab{} & = & 0\, . \label{PsWiv} 
\end{eqnarray}
Note, these are not Wronskian relations.
\begin{lem}
The matrix $\Mb{}{}(k)$ is invertable for finite real $k$.
\end{lem}
{\it Proof:}
We claim that for real $k$
$$
\Mb{}{} L^{*} + L \Mb{*}{} = \1
$$
where $L=\Jfa{,x} B + \Jfa{} A$. Since $k$ is real
$\Mb{*}{}(k)=\Ma{}{}(k)$ and similarly $L^{*}=B^{*} \Jfb{,x} +
A^{*} \Jfb{}$ so the left hand side becomes
$$
\mbox{} -\frac{1}{2ik} \left[ \left[ \Jfa{} B - \Jfa{,x} A \right] 
\left[ B^{*} \Jfb{,x} + A^{*} \Jfb{} \right] +
\left[ \Jfa{,x} B + \Jfa{} A \right]
\left[ A^{*} \Jfb{,x} - B^{*} \Jfb{} \right] \right] \, .
$$
Expanding this out and using $A A^{*} + B B^{*} =  \1$, $A B^{*} = B A^{*}$ and the Wronskian relations we immediately get the claimed equality. \\
Suppose there is a non-zero $a\in\ker\left( \Mb{}{}\left(\hat{k}\right) \right)$ for some $\hat{k}\in\R$. We have already shown that 
$$
\lim_{k\rightarrow\hat{k}}\left[ a^{*} \Mb{}{}(k) L^{*}(k) a + a^{*} L(k)
\Mb{*}{}(k) a \right] = a^{*}a \ne 0\, ,
$$
which can only hold if $L(k)$ has a pole at $\hat{k}$. But this supplies
a contradiction since, by theorem \ref{thJsi},
the elements of the matrix $L$ are bounded continuous functions for real
finite values of $k$. \hspace*{\fill} $\Box$ \\ 

This generalises lemmata 2.2.2 and 2.4.1 of \cite{Agr:Mar} where the result is proved for Dirichlet boundary conditions, ie $-2k\Md{}{}=\pm\Jfad{}$.  Consequently, for real $k$ we define the scattered wave $\Sw (x,k)$ and scattering matrix $S(k)$ in terms of $\eSw (x,k)$
$$
\Sw (x,k) \equiv \eSw (x,k) \Mb{-1}{} (k)= \Jsb{}(x,k) + \Js{}(x,k) S(k)
$$
where 
\begin{equation}\label{smii}
S(k) = - \left[ \Jfab{} B - \Jfab{,x} A \right] 
\left[ \Jfa{} B - \Jfa{,x} A \right]^{-1} \, . 
\end{equation}
It is clear from the definition that the scattering matrix can be extended off the real axis as a meromorphic function and we show that away from any poles
$$
S^{\dagger}=S^{-1}\, .
$$
Taking the Wronskian of $\eSwa$ and $\eSw$
$$
W\{\eSwa ,\eSw \} = A^{\star} B - B^{\star} A = 0 \, ,
$$
we see that it is zero. Furthermore, where the inverse $\Mb{-1}{}$
exists we can write $\eSw$ in terms of the scattering wave
\begin{eqnarray*}
W\{\eSwa ,\eSw \} & = & \Ma{}{} W\{\Swa ,\Sw \} \Mb{}{} \\
& = & 2ik \Ma{}{} \left[ -\1 + S^{\dagger} S \right] \Mb{}{} = 0 
\end{eqnarray*}
to get our result. In particular this gives the unitarity of the scattering matrix on the real axis. The high energy asymptotic of the scattering matrix is related to the selfadjoint boundary conditions by the following lemma:
\begin{lem}\label{umhu}
Given the self-adjoint $\cL$ with associated unitary matrix $U$
defining the boundary conditions the scattering matrix 
has the asymptotic
$$
\lim_{k\rightarrow\infty} S(k) = \hat{U}
$$
where $\hat{U}$ is a unitary hermitian matrix derived from $U$ by applying the map
$$
z\mapsto \left\{ \begin{array}{cl}
 1 & : z\in\T \setminus \{-1\} \\
-1 & : z = -1
\end{array}  \right.
$$
to the spectrum of $U$.
\end{lem}
Here $\T$ is the unit circle in $\C$. The proof follows from diagonalising $U$ and the asymptotics of the Jost solutions. \\
Our assumption that there are no virtual levels gives a simple proof of the fact that there are a finite number of discrete eigenvalues (see \cite{Agr:Mar} pg. 38) although this statement is still true when there are virtual levels (as may be proved using Glazman's method of splitting, see \cite{Agr:Mar} theorem 2.1.1 for the Dirichlet case). Consequently, we have a finite number of negative (this follows from the existence of an integral equation representation of the Jost solutions which appears in the proof of therorem \ref{thJsi}) discrete eigenvalues $k^2_l$. We write $k_l =i\kappa_l$ and choose the root $\kappa_l > 0$. The condition for a discrete eigenvalue is that, for some vector $a$, the square integrable vector function $\Js{}(x,k_l) a$ satisfies the self-adjoint boundary conditions (\ref{basicbc}) 
$$
\left[ \frac{i}{2}(U^{\star}-\1 ) \Jf{}(k_l) + \frac{1}{2}(U^{\star} +\1 ) \Jf{,x}(k_l) \right] a = 0 \, .
$$
This is equivalent to $\det (\Mab{}{}(k_l)) = 0$ and, since $k_l$ is purely imaginary and $\Mab{}{}(k_l)=-\Mb{*}{}(k_l)$ we can write this as $\det (\Mb{}{}(k_l)) = 0$. This gives:
\begin{thm}
The discrete eigenvalues of the self-adjoint Schr\"{o}dinger operator correspond to the zeroes of 
$$
\det (\Mb{}{}(k))
$$
in the upper half-plane.
\end{thm}
\begin{thm}\label{smpl}
The poles of $\Mb{-1}{}(k)$ in the half-plane $\Im (k) > 0$ are simple.
\end{thm}
We say that a matrix has a simple pole at $k_l$ if it can be expanded as a power series in $k-k_l$ with the lowest order term $(k-k_l)^{-1}\Nb{,l}$. \\
{\it Proof:} Let us consider the entire solution $\eSw$ of the eigenvalue equation (\ref{be}) and the solution $\eSwa$ of the `adjoint' equation. We differentiate this adjoint equation with respect to $k$, multiply on the right by $\eSw$, and subtract from it (\ref{be}) premultiplied by $\eSwa_{k}$. This gives
$$
\eSwa_{k} \eSw_{xx} - \eSwa_{xxk} \eSw = 2k \eSwa \eSw\, .
$$
We integrate the space variable from $x$ to $N$ and put $k=k_l$ to get 
\begin{equation}\label{qfrm}
\left. a^{*}\left[ \eSwa_{k} \eSw_{x} - \eSwa_{xk} \eSw \right] a
\right|^N_x = 2k_l \int^N_x [\eSw a]^{*} \eSw a \, dt 
\end{equation}
where $a\in\ker\left(\Mb{}{}(i\kappa_l)\right)$ is non zero. Using the fact that $a^{*}$ eliminates $\Mab{}{}$, $a$ eliminates $\Mb{}{}$ and the constancy of the Wronskian the left hand side simplifies to
$$
\left. a^{\star}\Ma{}{}\left[\Jsa{,k} \Js{,x} - \Jsa{,xk}\Js{} \right]
\M{}{} a \right|^N_x \, .
$$
Since $\Im (k_l)>0$ all of the terms in the bracket are exponentially
decreasing as $N\rightarrow\infty$ so the upper limit vanishes leaving
$$
\mbox{} - a^{\star} \Ma{}{} \left[\Jfa{,k} \Jf{,x} - \Jfa{,xk} \Jf{}\right]\M{}{}
a  = 2 k_l \int^{\infty}_0 [\eSw a]^{\star} \eSw a\, dt \neq 0 \, .
$$
We now expand out $\M{}{}$ in terms of Jost functions and use the identities (\ref{PsWi}-\ref{PsWiv}) and $\Mb{}{} a=0$ to get 
$$
i a^{\star}\Ma{}{} \Mb{}{,k} a = \int^{\infty}_0 [\eSw a]^{\star} 
\eSw a\, dt \neq 0 \, .
$$
Now it is well known that $\Mb{-1}{}(k)$ has a simple pole at $k=k_l$ iff the relations
\begin{eqnarray*}
\Mb{}{}(k_l) a & = & 0 \\
\Mb{}{}(k_l) b + \Mb{}{,k}(k_l) a & = & 0 
\end{eqnarray*}
for some $b$ implies that $a=0$. Premultiplying the second
relation by $a^{\star}\Ma{}{}$ gives 
$$
a^{\star}\Ma{}{} \Mb{}{} b + a^{\star}\Ma{}{} \Mb{}{,k} a = 0 
$$
and it is easy to see that the first term is zero: we use (\ref{PsWi}-\ref{PsWiv}) and the fact that $a^{*}$ eliminates $\Mab{}{}$. But this implies that $a^{\star}\Ma{}{} \Mb{}{,k} a = 0$ which can only be true if $a=0$. This implies that the pole is simple. \hspace*{\fill} $\Box$ 
\section{The Inverse Scattering Problem}
In this section we derive the inverse scattering problem as an integral equation problem---the Marchenko equation (a discussion using the Riemann Hilbert problem is given in \cite{Har}). Our solution is to some degree formal; to treat this problem in its entirety we should first give a complete description of the space in which the scattering data exists and then show that the inverse scattering problem has a solution for any element of this space. A complete description of the space of scattering data for the matrix Scr\"{o}dinger operator with general boundary conditions does not (to the knowledge of the author) exist. Indeed this space would almost certainly depend on the boundary condition $U$ (or what seems likely $\hat{U}$). \\
Nevertheless, given the condition (\ref{bcndxx}) as well as the assumption of no virtual levels it is possible to show that the (scattering) data $S(k)-\hat{U}$ is the Fourier transform of a hermitian matrix with integrable entries. This important result follows from theorem E.0.3 of \cite{Har}. We then have from theorem 3.4.1 of \cite{Agr:Mar} that the Marchenko equation has a unique solution, again with integrable entries. We simply assume here a very narrow class of operators (certainly narrower than in \cite{Agr:Mar}) but a class for which it is easy to show that the inverse problem has a solution. 
\subsection{The Marchenko Equation}
The novel feature of the inverse problem for general boundary conditions is that the boundary conditions appear in the scattering data and the inverse problem through the asymptotic $\hat{U}$. \\
We start the derivation of the Marchenko equation by substituting (\ref{beJT}) into $\Sw - \Js{} \hat{U}$
\begin{eqnarray*}
\Sw - \Js{} \hat{U} & = & e^{-ikx}\1 + e^{ikx} (S - \hat{U}) + \int^{\infty}_{x} K(x,t)
 e^{-ikt}\1\, dt \\
& & \mbox{} + \int^{\infty}_{x} K(x,t) e^{ikt} (S - \hat{U})\, dt \, .
\end{eqnarray*}
As noted above the Fourier transform
$$
G_c (y) \equiv \frac{1}{2\pi} \int^{\infty}_{-\infty} \left(S (k) - \hat{U}\right)
e^{iky} \, dk
$$
exists (in fact is hermitian and integrable) so
\begin{eqnarray}
\frac{1}{2\pi} \int^{\infty}_{-\infty} \left( \Sw (x,k) - \Js{}(x,k)
\hat{U}\right) e^{iky} \, dk & = & G_c(x+y) + K(x,y) +  \mbox{} \nonumber \\
& & \hspace*{-8mm}\mbox{} + \int^{\infty}_{x} K(x,t) G_c(t+y)\, dt \label{preMe}
\end{eqnarray}
where we will only consider $x<y$. The left hand side is exponentially decreasing in the upper half plane (since $x<y$) and therefore we can close the contour of integration there. Since, theorem \ref{smpl}, the scattered wave has only simple poles at the eigenvalues there appears a sum of residues on the left hand side
\begin{eqnarray*}
\frac{1}{2\pi} \int^{\infty}_{-\infty} \left( \Sw (x,k) - \Js{}(x,k)
\hat{U}\right) e^{iky} \, dk  & = & \frac{1}{2\pi} \int^{\infty}_{-\infty} \eSw (x,k) \Mb{-1}{} e^{iky}\, dk \\ 
& = & i \sum^{N}_{l=1} \eSw (x,k_l) \Nb{,l}\, e^{ik_l y}
\end{eqnarray*}
where $\Nb{,l}$ is the residue of $\Mb{-1}{}$ at the eigenvalue $k=k_l$. As $\Js{}$ is analytic in the upper half plane it makes no contribution. \\
To simplify the form of the residues we first need two auxiliary results. Defining $P_l$ to be the orthogonal projection onto $\ker \Mab{}{}(k_l)$ we have
\begin{equation}\label{eswJs}
\eSw (x,k_l) U^{\star} ( \Jf{} - i \Jf{,x} )(k_l) P_l = \Js{}
(x,k_l) P_l \, .
\end{equation}
This just follows if we evaluate the left hand side at $x=0$, and the $x$ derivative of the left hand side at $x=0$ and see that we get the right hand side and its $x$ derivative at $x=0$. The second result
\begin{equation}\label{IPz}
\Mb{}{} U^{\star} ( \Jf{} - i \Jf{,x} )(k_l) P_l = 0 
\end{equation}
follows if we expand $\Mb{}{}$ and use the left hand side of (\ref{eswJs}), again evaluated at $x=0$, to get the Wronskian $W\{ \Jsa{}, \Js{} \}$ which is zero. \\
Let us consider the power series of $\Mab{}{}(k_l) = - \Mb{*}{}(k_l)$ and its inverse
\begin{eqnarray*}
\Mab{}{} (k) & = & \Mab{}{}(k_l) + (k-k_l) \Mab{}{,k}(k_l) + \cdots  \\
\Mab{-1}{}(k) & = & (k-k_l)^{-1} \Nab{,l} + \Oab{,l} + \cdots
\end{eqnarray*}
where the subscript $k$ denotes differentiation with respect
to $k$ and the subscript $l$ is an index of the zeroes $k_l$. These
expansions give us the following relations
\begin{eqnarray}
& \Mab{}{}(k_l) \Nab{,l} = \Nab{,l} \Mab{}{}(k_l) = 0 & \label{resreli} \\
& \Mab{}{,k}(k_l) \Nab{,l} + \Mab{}{}(k_l) \Oab{,l} = \Nab{,l}
\Mab{}{,k} (k_l) + \Oab{,l} \Mab{}{}(k_l)  = \1 \, . & \label{resrelii}
\end{eqnarray}
From (\ref{resreli}) we have $P_l \Nab{,l} = \Nab{,l}$ or taking the complex conjugate transpose
\begin{equation}\label{resprj}
\Nb{,l} P_l = \Nb{,l} \, . 
\end{equation}
We will now show that the left hand side of (\ref{eswJs}) is `close to' the residue, namely
\begin{equation}\label{eswJsi}
U^{\star}( \Jf{} - i \Jf{,x} ) P_l = i \Nb{,l} A_l P_l
\end{equation}
where $A_l$ is the positive definite hermitian matrix
$$
A_l\equiv \int^{\infty}_0 \Js{}^{\star} \Js{} (t,k_l) \, dt \, .
$$
To see this we take the left hand side of (\ref{eswJsi}), multiply on the left by the complex conjugate transpose of equation (\ref{resrelii}), expand out and use (\ref{IPz}), then $\Mab{}{}(k_l) P_l = 0$ to get
$$
U^{\star}( \Jf{} - i \Jf{,x} ) P_l = - \frac{1}{2ik_l} \Nb{,l} \left[ \Jfa{,k}\Jf{,x} - \Jfa{,xk}\Jf{} 
\right] P_l \, .
$$
Using the same reasoning as was used to derive (\ref{qfrm}) we see that the term in brackets $\Jfa{,k}\Jf{,x} - \Jfa{,xk}\Jf{} = 2 k_l A_l$ which gives (\ref{eswJsi}). \\
Going back to (\ref{eswJs}) we get
$$
\Js{} (x,k_l) P_l = i\, \eSw (x,k_l) \Nb{,l} A_l P_l  = i\, \eSw (x,k_l) \Nb{,l} B_l 
$$
where $B_l = P_l A_l P_l + P^{\perp}_l$ and we have used (\ref{resprj}). It is clear that $B_l$ is positive definite so we can define 
$$
C_l\equiv P_l B^{-1/2}_l \, . 
$$
Then $C^2_l = P_l B^{-1}_l  $ from which we finally get the desired form for the residue
\begin{equation}\label{degf}
\Js{} (x,k_l) C^2_l = i\, \eSw (x,k_l) \Nb{,l} \, .
\end{equation}
The $C_l$ are known as normalisation matrices as the columns of $\Js{}(x,k_l)C_l$ form a complete set of normalised eigenfunctions (this is a simple consequence of the above, for details see \cite{Agr:Mar}).
From (\ref{beJT}) and (\ref{preMe}) we immediately get the Marchenko equation (\ref{Me}).
\begin{thm}
Given the scattering data
$$
\{ S(k); \: \kappa_l, \: C_l, \: l=1,\ldots ,N \}
$$
where $S(k)$ is a unitary matrix, $C_l$ are non-negative hermitian matrices and $\kappa_l$ are positive real numbers we can recover the potential of the matrix Schr\"{o}dinger operator from the solution of the Marchenko equation
\begin{equation}\label{Me}
G(x+y) + K(x,y) + \int^{\infty}_{x} K(x,t) G(t+y)dt = 0 \hspace{5mm} x<y,
\end{equation}
where
$$
G(t) = \sum^{N}_{l=1} C^2_l e^{-\kappa_l t} + \frac{1}{2\pi}
\int^{\infty}_{-\infty} (S (k) - \hat{U}) e^{ikt} dk \, .
$$
Here $\hat{U}=\lim_{k\to\infty}S(k)$. 
\end{thm}
{\it Proof:} The potential is recovered from the solution of the Marchenko equation using (\ref{beIP}). \hspace*{\fill} $\Box$ \\
\begin{cor}
Given the scattering data and the solution $K(x,t)$ of the Marchenko equation we can recover the selfadjoint boundary conditions of the matrix Schr\"{o}dinger operator from
$$
U = \left( \left.\Sw\right|_0 - \left.i \Sw_x\right|_0\right) \left( \left.\Sw\right|_0
+ \left.i \Sw_x\right|_0 \right)^{-1} \, .
$$
\end{cor}
{\it Proof:} This follows from (\ref{eSwo}). The scattered wave can be found from the scattering matrix and (\ref{beJT}). \hspace*{\fill} $\Box$ \\
\subsection{The Diagonal Potential}
We finish with some brief comments on the case where $Q(x)$ is a real diagonal matrix. This is easily seen to be equivalent to the scattering problem for a star-shaped graph with a single node and $n$ semi infinite rays. In this case we expect some simplification to occur:
\begin{prop}
In the case of a diagonal potential the
following scattering data is sufficient to recover the potential
$$
\{ R_i(k); \: \kappa_l, \: \gamma_{l,i}, \: i=1,\ldots ,n  \; l=1,\ldots ,N\} 
$$
where $R_i(k)\equiv S_{ii}(k)$, known as the reflection coefficients, are
the diagonal elements of the scattering matrix and $\gamma_{l,i} \equiv C^2_{l,ii}$ are the diagonal elements of the squares of the normalisation
matrices, known as the normalisation constants.
\end{prop}
{\it Proof:} In the case of a diagonal potential the
kernel of the transformation operator $K(x,t)$ is, like the
Jost solution $\Js{}$, a diagonal matrix. Consequently the diagonal elements of the 
Marchenko equation (\ref{Me}) form $n$ independent scalar Marchenko equations. It is easy to see that these scalar Marchenko  equations can be solved using only the above
scattering data. \hspace*{\fill} $\Box$ \\

There is good reason to expect that we can do better than this, but only if we are {\em a priori} given the form of the selfadjoint boundary conditions at the origin. We discuss this point for so called `flux conserved' boundary conditions in the publication \cite{Har1} and show there that it is possible to recover the potential with only $n-1$ reflection coefficients and normalisation constants.
\section*{Acknowledgements}
The author would like to thank Prof B. Pavlov for assistance and advice.


\end{document}